\documentclass[11pt,twoside]{article}
\usepackage{asp2010}
\usepackage{bm}

\resetcounters

\begin{document}

\title{Post-Newtonian prediction for the (2,2) mode of the gravitational wave emitted by compact binaries}

\author{Guillaume Faye,$^1$ Sylvain Marsat,$^1$ Luc Blanchet,$^1$ and Bala R. Iyer$^{2}$
\affil{$^1$Institut d'Astrophysique de Paris --- UMR 7095 du CNRS, Universit\'e
  Pierre \& Marie Curie, 98$^{\mathrm{bis}}$ boulevard Arago, 75014 Paris,
  France}
\affil{$^2$Raman Research Institute, Bangalore 560 080, India}}

\begin{abstract}
	We present our 3.5PN computation of the (2,2) mode of the gravitational wave amplitude emitted by compact binaries, on quasi-circular orbits and in the absence of spins. We use the multipolar post-Newtonian wave generation formalism, extending and building on previous works which computed the 3PN order. This calculation required the extension of the multipolar post-Minkowskian algorithm, as well as the computation of the mass source quadrupole at 3.5PN order. Our result will allow more accurate comparisons to numerical relativity, and is a first step towards the computation of the full 3.5PN waveform amplitude, which would improve the estimation of the source parameters by gravitational wave detectors.
\end{abstract}


\section{Introduction}\label{sec:intro}

Gravitational wave detectors such as VIRGO, LIGO and the planned space-based detector LISA, need an accurate theoretical prediction for the expected signal of compact binary systems. Cross-correlation with a bank of templates will allow on-line detection and off-line parameter estimation of the source. This has driven a lot of effort for gaining accuracy in our modelling of the signal, both with analytical and numerical methods. Templates currently in use combine post-Newtonian information for the inspiral phase \citep{Bliving} with numerical relativity results for the plunge, merger and ringdown phases~\citep{Pretorius05,BCCKvM06,CLMZ06}, which are matched together to provide the full gravitational waveform, e.g. \citet{PBFRT11}.

In the current state of the art of the post-Newtonian framework, the equations of motion and phase evolution of the binary are known up to the 3.5PN order, while the full waveform is known up to the 3PN order \citep{BFIS08} --- with as usual 1PN corresponding to $1/c^{2}$. Extending our knowledge of the waveform up to the 3.5PN order would improve the parameter estimation of the source. In particular, for supermassive black holes in the LISA range, it has been shown by \citet{AISSV07} and \citet{TS08} that a significant improvement on the angular resolution and distance measurement was obtained by including higher-order harmonics in the signal. Another interesting outcome of higher-order computations will be to compare the results obtained with the ones of the numerical relativity codes.

In this work which is a summary of \citet{FMBI12}, we address the computation of the 3.5PN contribution to the quadrupolar mode (2,2) of the waveform. This mode corresponds to the leading order quadrupolar radiation, and it is the mode which is computed the most accurately by numerical codes. Our result constitutes a first step towards the computation of the full 3.5PN waveform.


\section{3.5PN equations of motion in the quasi-circular case}\label{sec:eom}

While addressing the computation of the 3.5PN order of the quadrupolar mode in the waveform, we shall need 3.5PN order information on the equations of motion as well, which is known from previous works. We model the inspiralling compact binary by two point-like particles characterized by their masses $m_1$ and $m_2$. Denoting the relative position and velocities of the two particles by $\bm{x} = \bm{y}_{1} - \bm{y}_{2} = r \, \bm{n}$ and $\bm{v} = \mathrm{d}\bm{x}\mathrm{d} t$, using the notation $\bm{\ell} =\bm{x}\times \bm{v} / \vert \bm{x} \times \bm{v}\vert$ for the unit vector orthogonal to the orbital plane (which is fixed in the non-spinning case) and finally completing the triad by $\mbox{\boldmath$\lambda$} = \bm{\ell}\times\bm{n}$, we have for quasi-circular orbits
\begin{eqnarray}
	\bm{v} &=& \dot r \,\bm{n} + r \,\omega \,\mbox{\boldmath$\lambda$}\,,\label{eq:v}\\ 
	\frac{\mathrm{d} \bm{v}}{\mathrm{d} t} &=& - \omega^2 \,\bm{x} + \Bigl(\frac{\dot{\omega}}{\omega} + 2\frac{\dot{r}}{r}\Bigr)\,\bm{v}\,+ \mathcal{O} \left(10\right)\,. \label{eq:a}
\end{eqnarray}
Here $\mathcal{O}(n)$ means $\mathcal{O}(c^{-n})$, and the orbital frequency is $\omega=\dot{\varphi}$ with $\varphi$ being the orbital phase. We used the facts that $\dot{r}=\mathcal{O}(5)$, $\ddot{r}=\mathcal{O}(10)$,  $\dot{\omega}=\mathcal{O}(5)$ for a quasi-circular orbit decaying by the effect of gravitational radiation.

On one hand, the information about the conservative part of the dynamics for circular orbits is contained in the generalized Kepler's law $\omega(r)$. This relation has been computed in harmonic coordinates~\citep{BDE04}, in Arnowitt-Deser-Misner (ADM) coordinates \citep{DJSdim}, and recently retrieved by effective field theory methods \citep{FS11}. It reads, defining $m \equiv m_1+m_2$, $\nu=m_1m_2/m^2$, and the PN parameter $\gamma \equiv G m / r c^2$~:
\begin{eqnarray}\label{eq:omega3PN}
\omega^2 &=& {G m\over r^3}\left\{ 1+\gamma\Bigl(-3+\nu\Bigr) + \gamma^2 \left(6+\frac{41}{4}\nu +\nu^2\right) \right. \nonumber\\
&&\left. +\gamma^3 \left(-10+\left[-\frac{75707}{840}+\frac{41}{64}\pi^2 +22\ln\left(\frac{r}{r'_0}\right) \right]\nu +\frac{19}{2}\nu^2+\nu^3\right)\right\} +\mathcal{O}(8) \,.
\end{eqnarray}
The constant $r'_0$ is specific to harmonic coordinates and disappears from the physical results in the end. On the other hand, the 2.5PN and 3.5PN contributions to the equations of motion correspond to radiation reaction which drives the evolution of the orbital parameters according to~:
\begin{eqnarray}\label{eq:rdot}
	\dot{r} &=& - \frac{64}{5} \sqrt{\frac{G m}{r}}~\nu\,\gamma^{5/2}\left[1+ \gamma \left(-\frac{1751}{336} - \frac{7}{4}\nu\right)\right] + \mathcal{O}(8)\,, \nonumber \\
  	\dot{\omega} &=&
\frac{96}{5} \,\frac{G m}{r^3}\,\nu\,\gamma^{5/2}\left[1+ \gamma \left(-\frac{2591}{336} - \frac{11}{12}\nu\right)\right] + \mathcal{O}(8) \,.
\end{eqnarray}
%


\section{Gravitational waveform for non-spinning binaries}\label{sec:gwf}

\subsection{Definitions and notations}

Defining an asymptotic Bondi-type coordinate system $X^\mu=(c T,\bm{X})$ \citep{Th80}, the gravitational waveform may be parametrized by so-called radiative symmetric, trace-free (STF) mass and current multipole moments $U_{L}$ and $V_{L}$ according to 
\begin{eqnarray}\label{eq:hij}
	h_{ij}^\mathrm{TT} &=& \frac{4G}{c^2R} \,\mathcal{P}_{ijkl}
(\bm{N}) \sum^{+\infty}_{\ell=2}\frac{1}{c^\ell \ell !} \biggl\{ N_{L-2}
\,U_{klL-2}(T_R) \nonumber \\
	&& \qquad \qquad \qquad \qquad - \frac{2\ell}{c(\ell+1)} \,N_{aL-2} \,\varepsilon_{ab(k}
\,V_{l)bL-2}(T_R)\biggr\} \,,
\end{eqnarray}
keeping only terms of order $\mathcal{O}(1/R)$, with $R$ and $\bm{N}$ the radial distance and unit vector, and $T_R=T-R/c$ the retarded time. We use a multi-indices notation $A_{L}=A_{i_{1}\dots i_{l}}$, and $\mathcal{P}_{ijkl}$ is the transverse-traceless projector. Next, the two polarizations of the waveform are defined as
\begin{eqnarray}
h_+ &=& \frac{1}{2}\left(P_iP_j-Q_iQ_j\right) h_{ij}^\mathrm{TT}\,,\\
h_\times &=& \frac{1}{2}\left(P_iQ_j+P_jQ_i\right) h_{ij}^\mathrm{TT}\,,
\end{eqnarray}
where we take as a convention for the polarization vectors $\bm{P}=-\bm{e}_{\Phi}$ and  $\bm{Q}=\bm{e}_\Theta$, where $(\Theta,\Phi)$ refer to the direction $\bm{N}$ towards the observer. Decomposing $h \equiv h_+ -\mathrm{i} h_\times$ in terms of spin-weighted spherical harmonics of weight $-2$, we define the spherical modes of the waveform $h^{\ell m}$ by~:
\begin{equation}\label{eq:hdecomp}
	h = \sum^{+\infty}_{\ell=2}\sum^{\ell}_{m=-\ell} h^{\ell m} \,Y^{\ell m}_{-2}(\Theta,\Phi)\,.
\end{equation}
This defines in particular the $h^{22}$ mode of the waveform, the computation of which we address in this work. Notice that, from the rotational invariance of the system, it can be shown that $h^{\ell m}$ depends on the orbital phase $\varphi$ as $e^{-i m \varphi}$, so that $h^{22}$ corresponds to the quadrupolar radiation at twice the orbital frequency.

\subsection{Mode separation for planar binaries}

Now, let us stress an important property, valid for any binary without precession of the orbital plane, possibly on an elliptical orbit or including spins aligned or anti-aligned with the orbital angular momentum. The parity invariance of the problem  and the symmetry of revolution together translate into the following relations for the waveform~:
\begin{eqnarray} \label{eq:parity_hTT}
	h^\mathrm{TT}_{ij}(-\bm{X}, T; -\bm{y}_A, -\bm{v}_A) &=& h^{\mathrm{TT}}_{ij}(\bm{X}, T; \bm{y}_A, \bm{v}_A)\, , \\
	h(\pi-\Theta, \phi) &=& \overline{h}(\Theta, \phi) \, ,
\end{eqnarray} 
where $\bm{X}$ is the position of the observer, with angular coordinates $(\Theta,\Phi)$, and $\bm{y}_A, \bm{v}_A$ are the positions and velocities of the two massive bodies. We set $\phi \equiv \varphi - \Phi + \pi/2$, so that $h$ depends only on $\Theta$ and this relative phase $\phi$ by rotational invariance.

Using the properties of spin-weighted tensor spherical harmonics, one obtains~: 
\begin{eqnarray} 
	h^{\ell m} &\propto & U_{L}\quad \mathrm{when} \quad \ell+m \quad \mathrm{even} \, , \nonumber \\
	h^{\ell m} &\propto & V_{L}\quad \mathrm{when} \quad \ell+m \quad \mathrm{odd} \, .
\end{eqnarray} 
In particular, $h^{22}$ depends only on $U_{ij}$ and not on $V_{ij}$, and this dependence reads (with $1,2,3$ labelling the Cartesian coordinates, and recalling that $U_{ij}$ is STF)~:
\begin{equation}\label{eq:h22}
	h^{22} = - \frac{G}{R c^4} \sqrt{\frac{4\pi}{5}} \Bigl(\delta_{i}^1-\mathrm{i}\delta_{i}^2\Bigr)\Bigl(\delta_{j}^1-\mathrm{i}\delta_{j}^2\Bigr) \, U_{ij} \,.
\end{equation}
%


\section{Computation of the quadrupole mode}\label{sec:quadmode}

As we have seen, the computation of the $h^{22}$ mode reduces to the one of the radiative mass quadrupole $U_{ij}$. Extending the multipolar-post-Minkowskian (MPM) formalism~\citep{BD86,B98mult,BFIS08} to 3.5PN order, we express $U_{ij}$ first in terms of canonical moments $\{M_L, S_L\}$, and then of source moments $\{I_L, J_L\}$ and gauge moments $\{W_L, X_L, Y_L, Z_L\}$ which correspond to higher-order corrections. The result of this algorithm reads schematically (the explicit formulas are too long to be displayed here) :
\begin{equation}\label{eq:U2decomp}
U_{ij} = U^\mathrm{inst}_{ij} + U^\mathrm{tail}_{ij} + U^\mathrm{mem}_{ij}\,.
\end{equation}
We find that no cubic interactions between momenta arise at this order, for the radiative mass quadrupole. $U^\mathrm{inst}_{ij}$ contains the leading order contribution, which is simply the second time derivative $M_{ij}^{(2)}$, as well as higher-order instantaneous interactions between the $M_{L}$ and $S_{L}$. The tail and memory contributions $U^\mathrm{tail}_{ij}$ and $U^\mathrm{mem}_{ij}$ are hereditary, and take the form of integrals over the whole past of the source, with a logarithmic kernel in the case of tails. Then, the canonical moments $\{M_L, S_L\}$ can be reexpressed in terms of the source and gauge moments $\{I_L,
J_L, W_L, X_L, Y_L, Z_L\}$, which themselves admit expressions as integrals, over space, of the stress-energy pseudotensor which includes both matter and non-linearities in the metric perturbation. At leading order, canonical and source moments are identical, and they start differ at 2.5PN order.
 
Along with lower-order expressions for the other moments, the quadrupole source moment $I_{ij}$ has to be computed to the 3.5PN order. For quasi-circular orbits, it admits the following structure, with brackets for STF projection~:
\begin{equation}\label{eq:I3PN}
I_{ij} = \nu\,m\,\left(\mathcal{A}\,x^{\langle i}x^{j\rangle} +\frac{r^2}{c^2} \,\mathcal{B}\,v^{\langle i}v^{j\rangle} + \frac{48}{7}\frac{G^2 m^2 \nu}{c^5 r}\,\mathcal{C}_\mathrm{RR}\, x^{\langle i}v^{j\rangle}\right)+\mathcal{O}\left(8\right)\,.
\end{equation} 
The coefficients $\mathcal{A}$ and $\mathcal{B}$ correspond to the conservative part and are already known up to 3PN order \citep{BI04mult}. The 3.5PN contribution, which we had to compute, enters only the radiation reaction coefficient $\mathcal{C}_\mathrm{RR}$, for which we obtain~:
\begin{equation}\label{eq:C}
	\mathcal{C}_\mathrm{RR} = 1+\gamma\left(-\frac{256}{135}-\frac{1532}{405}\nu \right) + \mathcal{O}(3) \,.
\end{equation}

Notice finally that the assumption of a quasi-circular orbit plays a crucial role when computing hereditary integrals in the memory and tail terms.


\section{Result}\label{sec:h22}

We absorb logarithmic terms, originating from the tails, by the following change of phase variable~:
\begin{equation}\label{eq:changephaseomega}
\psi = \varphi - \frac{2 G M \omega}{c^3}\ln\left(\frac{\omega}{\omega_0}\right) \, ,
\end{equation}
which takes in fact the same form as at lower order, but where the ADM mass $M$ includes its 2PN corrections; $\omega_{0}$ relates to a constant $\tau_{0}$ introduced in the kernel of the tail integrals. Posing~:
\begin{equation}\label{eq:modedef}
  h^{\ell m} = \frac{2 G \,m \,\nu \,x}{R \,c^2} \,\sqrt{\frac{16\pi}{5}}\,H^{\ell m}\,e^{-\mathrm{i} m \, \psi} \,,
\end{equation}
with $x\equiv (G m \omega/c^{3})^{2/3}$, we finally obtain the following result for $H^{22}$~:
\begin{eqnarray} \label{eq:h22res}
H^{22} &=& 1+x \left(-\frac{107}{42}+\frac{55}{42}\nu\right)+2 \pi
  x^{3/2}+x^2
  \left(-\frac{2173}{1512}-\frac{1069}{216}\nu+\frac{2047}{1512}\nu^2\right)
  \nonumber \\ 
  &&+ x^{5/2} \left(-\frac{107 \pi }{21}-24 \,\mathrm{i}\,\nu +\frac{34
      \pi}{21}\nu\right)+x^3
  \bigg(\frac{27027409}{646800}-\frac{856}{105}\,\gamma_\mathrm{E} +\frac{428 
    \,\mathrm{i}\,\pi }{105}+\frac{2 \pi ^2}{3}\nonumber \\
  &&\qquad \qquad+ \left(-\frac{278185}{33264}+\frac{41 \pi^2}{96}\right) \nu
  -\frac{20261}{2772}\nu^2+\frac{114635}{99792}\nu^3-\frac{428}{105} \ln (16
  x)\bigg) 
  \nonumber \\
  &&+ x^{7/2} \left( -\frac{2173\pi}{756} + \left(
      -\frac{2495\pi}{378}+\frac{14333\,\mathrm{i}}{162} \right)\nu + \left(
      \frac{40\pi}{27}-\frac{4066\,\mathrm{i}}{945} \right)\nu^2 \right) \,,
\end{eqnarray}
where $\gamma_{E}$ is the Euler's constant appearing at 3PN. The newly computed 3.5PN terms in this expression are the ones proportional to $x^{7/2}$. We find agreement for $\nu\to 0$ with the test mass limit obtained by black-hole perturbation theory in \citet{TSasa94}.

This new result is ready for comparison with high-precision numerical relativity outputs. However, the computation of the other modes to the 3.5PN order is not straightforward, especially for the $(2,1)$ mode, and is left for future work.

\acknowledgments 
GF, LB, and BRI thank the Indo-French Collaboration (IFCPAR) under which
this work has been carried out.

\bibliography{marsat}

\end{document}